# Advancements in Biometric Technology with Artificial Intelligence

Konark Modi
Department of Computing and Informatics
Bournemouth University
Bournemouth, United Kingdom
s5515933@bournemouth.ac.uk

Lakshmipathi Devaraj
Department of Computing and Informatics
Bournemouth University
Bournemouth, United Kingdom
s5528240@bournemouth.ac.uk

*Abstract*—Authentication plays a significant part in dealing with security in public and private sectors such as healthcare systems, banking system, transportation system and law and security. Biometric technology has grown quickly recently, especially in the areas of artificial intelligence and identity. Formerly, authentication process has depended on security measures like passcodes, identity fobs, and fingerprints. On the other hand, as just a consequence of these precautions, theft has increased in frequency. In response, biometric security was created, in which the identification of a person is based on features derived from the physiological and behavioral traits of a human body using biometric system. Biometric technology gadgets are available to the public as they are embedded on computer systems, electronic devices, mobile phones, and other consumer electronics. As the fraudulent is increasing demand and use of biometric electronic devices has increased. As a consequence, it may be possible to confirm a person's distinct identification.

The goal of this study is to examine developments in biometric systems in the disciplines of medicine and engineering. The study will present the perspectives and different points of view of the secondary data, highlighting the need for more in-depth understanding and application of biometric technology to promote its development in the digital era. The study's findings may inspire people and businesses to more effectively incorporate biometric technologies in order to reduce the risks to data and identity security.

*Keywords—Advancements, biometric technology, artificial intelligence, authentication, security, physiological, behavioral., risks.*

## I. Introduction and background

A biometric system uses measurements of an individual's personal biological makeup or distinctive behaviors to identify them by comparison to biometric collection models of the same kind [1]. It is categorized as a recognition or validation system at the moment, and it can be described as a computerized method of determining or validating an individual's appearance based on specific physiological or behavioral characteristics [2].

Human society has depended on the capacity to distinguish between individuals and connect personal attributes (such as name, nationality, etc.) based on physical qualities like voice and appearance in addition to other context cues [3]. A person's character is composed of a variety of traits. In the early stages of human civilization, people dwelt in small communities where they could easily recognize each other. Nevertheless, in today's culture, the combination of greater mobility and fast population expansion has necessitated the creation of complex identity and access management systems that are capable of accurately capturing, preserving, and erasing individual identities. The need of having a reliable identity and access management system has been highlighted by the rising incidence of identity fraud and increasing national security concerns, which has led to biometric identification [4]. Identity management is essential in a wide range of applications, including trying to manage border checkpoints, restricting physiological access to vital facilities like power stations and airports, controlling logical access to public services and data, running virtual financial operations, and managing public assistance privileges. Due to the spread of web-based services and the establishment of decentralized client solution centers, identity fraud has escalated [5]

As stated by [6] the Federal Bureau of Investigation (FBI) started using suitable classification processes (invented by French criminologist Alphonse bertillon in 1879) to snap pictures of subject areas along with their height, each length, arm span, and pointer finger length in 1882, making biometric technology the earliest technique of and verify their identities. The development of speech, sign, handprint, and facial recognition software systems in addition to the involvement in biometrics shifting from straightforward hand metrics to eye character traits as well as other starting to emerge license plate recognition focusing on a wide range of bio - metric patterns (such as fingerprint, hand geometry, iris, and retina) during the 1990s. As research advances, several fresh, cutting-edge approaches to biometrics assessment are being investigated, such as the individual ear's shape, deoxyribonucleic acid (DNA), keystroke velocity, and body odor.

In contrast to the traditional technique of identifying, [7] define biometric as a distinguishing feature of an individual that does not vary over time, for instance a physiological or behavioral trait (e.g. pin, passwords). Handwriting signs and voice are examples of behavioral features, whereas physiological qualities include fingerprinting, iris, handprints, visage, and so forth. The words bio means life and metric means measuring, each of which have been in usage since the 20th century, are the inspiration for the names biometrics and biometry. Technological advancement uses the phrase biometric system to refer to systems that track human behavior and physical traits for purposes of verification and authentication [8]. The two parts of a biometric system are the enrollment module and the identification module. The enrollment component is responsible for instructing the system on how to identify a specific person by scanning their physiognomy and generating a digital representation of them. This electronic structure is used as a benchmark. The memory capabilities are in charge of recognizing a certain person by taking pictures of their characteristics and digitizing them into a format similar to the example. Recognition and verification are the two phases of the recognition process. During the define phase, the system asks, who is Jimmy? and tries to correlate Jimmy with each and every structure or characteristic in the database. On the other side, at the verification process, the system needs to make sure that the individual claiming to be Jimmy responds to the query "Is this Jimmy?" [9].

Biometrics can completely replace identity (ID) cards and credentials in worldwide surveillance systems. The common, originality, simplicity of acquisition, durability, mobility, and fakery tolerance of this method are all quite high. Biometrics is the measurement of an individual's physical, physiological, or behavioral characteristics, such as their face, voice, fingerprinting, handprints, iris and retinal scanners, biometric traits, and stance [10]. Traditional methods like pins, tokens, or credentials are unable to address the problem of being lost or stolen as there is an increasing demand in many modern security systems for safe access controls. In contrast, biometrics constitute a significant security improvement in addressing these new difficulties because they are based on "who" a person is or "how" they behave [11]. Consequently, biometrics offers important advantages over conventional approaches like tokens and credentials.

## II. STATEMENT OF THE PROBLEM

The most reliable form of authentication and verification has been through the use of biometric identification, an area that is quickly expanding. However, with the steadily increasing risks to information security, biometric has demonstrated a dependable secure communication but is still dealing with problems like the impact of a wet and creased fingerprint on the achievement of a biometric implementation is a problem that hasn't been thoroughly explored, corrective lenses and watery eyes in the particular instance of iris recognition, and so on [12]. It is commonly questioned where and to what extent biometric traits are distinctive and consistent among individuals. The developments in computer identity systems is not diverse enough [13], to enhance social processes sufficiently. As they can significantly aid in the construction of reliable biometric algorithms and in enhancing the usefulness and safety of biometrics, multi-classifier systems (MCS) and man-machine interface (MMI) demand extremely diversified interests [14].

In order to combat terrorism and identity fraud, more research and financing should indeed be promoted and permitted; in order to obtain the best degree of protection, multiple biometric methods should be supported [8]. As [15] points out, future analyses of artificial intelligence (AI) technology combined with authentication will likely enable massive increase because biometrics are a key component of AI technology. There is no recognized bio-metric trait that is totally continuous and unique because biometric systems comprise underlying data characteristics, uniqueness, and dynamic capabilities of stabilization under actual tangible conditions and ecologic hurdles, some of which are unidentified and treated, especially in a large number [16]. The core biological qualities and distribution of biometric data in a group are typically only discernible with filters enforced by biometric choice and measurement methodologies, which further complicates issues. In order to examine the potentials that can be suddenly realized from particular biometric technological applications, including such registered traveler operation, focuses on the identification, authentication method, and payment services in such sectors, there are also a few gaps in significant biometric systems that must be discussed in the medical and technological sectors [17]. As the interaction between the two can be of enormous importance to biometrics improvement to the engineering and medical industry, finding a good equilibrium between identification and safety precision has been a long-term area that has not yet been properly studied [18]. Given the limitations of accuracy, scale, security, and privacy—which have been the main categories of issues facing biometrics—biometrics is unique in that it involves judgment under uncertainty on both automatic speech recognition software and the subjective interpretation of its findings [4]. In order to speed up processes and procedures in both the engineering and medical industries, biometric technology is employed to produce more diagnostics and camera services like digital guardian sensors [19].

## III. REVIEW OF LITERATURE

It's important to consider several factors while selecting a biometric. It's crucial to comprehend the surroundings and requirements of biometrics. It is necessary to assess the level of success if a specific biometric is used for security purposes [20]. However, because no single biometric technology works well, a variety of biometric strategies are used to offer the highest level of security. A biometric authentication must take into account a number of factors, such as tasks, user circumstances, security concerns, current data, user count, etc [11]. If you've ever forgotten your computer or active notifications password, you can appreciate the significance of biometrics. In addition to rational usage, a fingertip scan or other kind of technology determines whether an individual is granted access to data or not, according to a 2007 report published by PBworks. The need to address user identity issues and lower passwords administration costs is growing. In higher education institutions, where passwords are tough for pupils to memorize, they routinely stole other students' login information and utilized it for their own purposes [21]. A biometric system is crucial to the existing level of protection in these situations.

The oldest biometric method that has been used successfully in a range of businesses is thumbprint identification. It is known that each person's fingerprint is unique. An imperfect imprint on the hand's surface is fingerprints. The arrangement of rough surfaces and the little spots aid in proving a fingerprint's individuality [21]. A scan that creates an image of the fingerprint is frequently used to gather fingerprints. The image of a fingerprint will then be examined using hybrid, minute details, or template methods [22]. Because fingerprint recognition is relatively remarkable in terms of its durability, ubiquity, accuracy, originality, and affordability, it is the most widely used and trusted method in addition to the most prevalent biometric technology [23]. They are widely used in forensics units, authorization entrance setup, in-network entry, and financial firms. Even though the biometrics site's recognition rate declines when the finger is wet and wrinkly. It has not yet been fully addressed how a wet and wrinkled finger affects a biometric device's performance. Additionally, [23] study found that noisy data can result in dust film deposition on a detector or from environmental factors, and that prints can be replicated in latex using an individual material.

The human palm can be used as a biometric attribute for authentication because its design comprises outcome based with unique measurements for each user and can't be altered. [7] Claim that two or three of the participant's fingers should be scanned in order to identify the user because this often requires little storage capacity. The palm print of a person's hand is photographed using a robust sensor module (optical reader), and kept as a pattern for comparing and distinguishing persons. Despite this, the massive number of data that was gathered prevents the system from being used in practice [2].

The tri-geometry of the hand is scanned by a scanning device to determine somebody's identification. This produces a numerical description that can then be matched to an image stored in an archive [8]. It suggests that modest security devices have a promising future. Hand recognition-based identification has quickly risen to the top of the biometric category due to its ease of learning, broad user acceptance, and dependability. Data sources based on digital webcams, electronic scanners, and camcorders are commonly used in offices and need less systems engineering labor. Palm prints are said to be more precise than fingerprinting since a body part is larger than a finger. Palm print images are resistant to fading and can be acquired using low-resolution lenses and scans while still maintaining enough info to identify high identification rates [23]. Despite this being true, obtaining photos in an uncontrolled environment with illumination changes and distortions due to human hand is also an issue. If the resolution of the photographic and camcorders used to take palm print shots is weak, Pictures may be difficult to notice. More research is required to determine how damp and wrinkled palms affect the rate of recognition [7,23].

There are various distinctive characteristics in each individual's iris that might be used to identify them. The iris' most prominent visible feature is tissue, which begins to grow constantly in every person at the young age of eight months [24]. In regards to user verification, iris images are taken by the detector and the iris constructions are evaluated using various iris dBs such as UPOL, MMU, IITD, and others. The iris is a hued heavily muscled circle from around the eye's center that includes the pupil dilation region and the cingulate region, and it is located by both the cornea and the lens of the eyeball [7]. There isn't any genetic impact on its development; instead, during the seventh months of pregnancy, a practice known as anarchic morphogenetic helps the development of the iris, enabling even identical twins to have separate irises [2]. According to [23] iris recognition uses video cameras and low-level infrared illumination to take pictures of the intricate, intricately detailed features of the iris without endangering or upsetting the subject. Digital templates created from these characteristics by computational and statistical algorithms make it feasible to identify a person or anyone purporting to be that person. The iris is a robust organ because it is inside.

This mode does not change as time passes from the first year after birth till death.

Each person has unique DNA, which is stable over the course of a lifetime. Deoxyribonucleic acid (DNA) is the most reliable type of biometric unique identifying method since the individual body is made up of around 60 trillion cells. Each body's nucleus is home to DNA, which is regarded of as the blueprint for the building of the female organism. DNA does not change during or after a person's life [25]. A sample of saliva, blood, sperm, hair, or tissues is needed for the labor-intensive authentication method of DNA recognition [2]. It may well be sustained positive to start a DNA-led identification process when getting recognition is difficult, particularly after a violent war. DNA analysis can produce a profile that can be confidently compared to other patterns. Inherently digital, DNA remains constant both during life and after death. Our physical and mental characteristics are determined by our foundation; unless someone is an exact duplicate, neither any person should share the same DNA gathering [23]. The difficulty with this could be that the pure compound needs to be enhanced and the DNA detection method must be improved. The length of DNA testing is the biggest problem. Since DNA matching requires complex chemical procedures that call for specialist skills where implementations are impossible [7].

Identification of keystroke dynamics with the use of biometrics allows for the pattern-based analysis of a person's keyboard behavior. To make the technology more durable and distinctive, the methods are always being refined. This system tracks keystroke acceleration and pressure, in addition to the total amount of time needed to enter the password and the intervals between punches on the keyboard. Access to technology might be advantageous since these biometrics may be used to verify the identity of the internet user continuously [21]. Keystroke recognition functions differently from the other biometric techniques that have been looked at. It is undoubtedly among the easiest to set up and run. This is because the present software-based approach to keyboard dynamics. All that is necessary is that the user use their current computer and keyboard [8]. It is possibly the only biometric technique that can be seamlessly integrated with open procedures and does not require the use of any additional, sophisticated equipment. It is also software-based [1]. Keystroke dynamics is still a recent approach that hasn't yet been deployed widely like other biometric systems are, though. Nevertheless, keystroke dynamics make access management and security on phones and computers straightforward. The researchers and developers still necessitate a standard important biometric repository, and the biometric authentication needs to be improved to maximize its efficacy [11].

Every person's face is distinctive, which is also a fundamental reality. The human face can be employed as biometrics to verify authentications. The idea behind biometric systems is to use facial features as a method of identification [26]. A face can be photographed and used as a comparison baseline thanks to good cameras. The framework is then contrasted using various methods of pattern matching to determine or confirm an individual's identification [12]. Numerous face datasets, including ORL, JAFFE, Yale, Multi

Pie, the AR data, the FERET database, and others are used to determine the rates [7]. Biometric technology is one of the most widely used biometric due to its simplicity and absence of distortion; it may be carried out using, among many other aspects, still pictures, video recording, stereoscopic, and ranging images. Despite the large number of examples in the galleries, face recognition under very collection conditions is more accurate and produces good identification rates [27]. In order to identify facial attributes, certain face recognition algorithms use feature or features from a subject's face image. For instance, an algorithm might consider the relative position, size, and/or shape of the eyes, nostrils, cheeks, and jaw. The next step is to look for a picture with the same qualities using these attributes. Using 3D face detection, a new development asserts to provide the highest levels of accuracy [15]. Whether the subject gives their approval or not, the photo is taken. Using well-designed equipment installed at airports, movie theaters, as well as other public locations, people can be recognized in crowds. Facial recognition software is also increasingly used to secure mobile phones [23]. Face recognition is a technology that the industry of mobile phones is testing with and putting into its devices.

This type of biometric focuses on the tone of the speaker's voice instead of trying to recognize words. This differs from technology that recognizes language and obeys commands. To avoid misinterpretation, the phrases "speaker recognition, confirmation, and identity" are employed [8]. The fundamental tenet of voice authentication is that every independent human voice can be identified by its unique pitch, tone, and volume. A cozy and consumer design are key technological problem in human-computer interaction. Because spoken languages predominate verbal discourse, people presume voice engagement with computers [23]. A previous sentence needs to be said in front of the sensor for voice recognition to function. The sensor will convert the audio information into a special digital code called a pattern, that will then be scrutinized to determine who the subject is. The tone of a voice signal is produced by reverberation in the larynx. The larynx's size in addition to the nose and mouth cavities' shapes have an impact on the sound that may be detected using this method [12]. Telephone apps are compatible with this biometric approach. The effectiveness of these gadgets, though, can be diminished by background noise in the area and phone system disruption.

An essential civic right, confidentiality underpins who we are in contemporary society and supports our ongoing struggle to maintain our autonomy in the midst of expanding state power. In order to safeguard the public interest and produce the right results for society, we must consider these rights when thinking of policy and regulation as a result of the "new technology realities" of today. Businesses, various government agencies, police departments, and other governmental and commercial entities are going to rely on bio - metric ultrasound more and more. These techniques must complement biometric confidentiality with a number of competing needs, including those of other people and western civilization as a whole [28]. It is hard to maintain the kinds of privacy people were accustomed to in the past since data is collected wherever we go and because of advancements in telecommunication, computers, and data collection, it is very simple to spread confidential details to anybody who is willing [8]. Biometrics may offer a quick and precise method of identification, enhancing security and secrecy, for instance by assisting an individual in preserving control over their data and reducing the likelihood of fraud. However, any use of biometric data raises questions regarding a person's ability to manage the information about themselves that they are willing to share with others, which could inevitably have an impact on their secrecy. Two factors of security concerns brought on by biometrics are individual rights (concerns over the loss of physical independence and human identity) and privacy protection (concerns over the misuse of data) [15]. If the verification program is written the person purposefully attempting to start exercising a selection to active participation in a system, the scheme does not necessitate the verification body to preserve vast amounts of information about a personal other than that is intended to prove that the person is legitimate, the use of biometrics for personal identification may present a lesser degree of data privacy risk. The lack of widespread adoption and consideration of biometric technology's privacy protections under current legislation creates a significant difficulty. Therefore, previous initiatives have provided some guidance for constructing the legal framework that would control future biometric technologies in order to address such legislative and regulatory issues. However, [29] states that suggestions should be made because different types of private information cannot be completely removed from biometrics authentication process, biometrics patterns, and the overall biometrics picture.

IV. METHODOLOGY

In order to evaluate the study's goal, the subjects' perceptions of biometric technology were examined using the qualitative research methodology, which is a systematic assessment of social practices in naturalistic environments [30]. Prior to beginning the qualitative approach, it was decided how much additional biometric technology could be developed based on prior studies. Semi-structured interviews, which were also utilized to gather information from participants in this study were employed to analyze participants' perspectives on biometric systems. Only volunteers from the medical and engineering departments with a professor of philosophy (Ph.D.) and master of science (M.Sc.) degree are expressly specified as requirements that each volunteer must meet in order to be eligible for the study. In the framework of evaluating qualitative data, the collected data will be analyzed using descriptive statistical analysis. Because it performs the same task as qualitative research in describing a situation and finding traits, frequencies, themes, and divisions, description was chosen (McCombes, 2019). The five phases of doing a descriptive statistical analysis framework, classifying data into the framework, analyzing the data, summarizing the findings, and present the results observed after the data were collected [31].

V. FINDINGS

Human physiological or behavioral traits can be measured using biometric technologies for security, identification, and verification. This study's biometrics component focuses on developments in biometric technologies. According to the data collected from those who participated in the medical and engineering fields, it appears that most attendees in the medical field are not familiar with the term "biometrics," but they become more cognizant after hearing a few interpretations they can understand, whereas respondents in the engineering field appear to know further about biometric systems. This might be the case because most technical experts will either have worked on developing biometric technology or researched the subject, or just because biometrics is more directly related to the technology age than the world of medicine. Because of this, engineering people are more knowledgeable and skilled about the subject than medical participants.

Most members from both areas agree that biometric technology is developing quickly in their respective fields. The research shows that biometric technology has been shown to be more effective and accurate for identity and data protection, which has increased interest inside the technology in the medical and engineering sectors as well as throughout the rest of the world. However, few medical professionals believe that biometrics is harmful to people's health, especially for those who already have certain medical issues. According to the participants, if biometric authentication is to be used in the industries, these risks cannot be completely eradicated but rather must be managed. Experts from the medical and technical fields claim that biometrics has more benefits than drawbacks since it offers better protection and is less dangerous. Biometric technology is an incomparably exceptional technology that possesses the capacity to transform the world in ways that are unimaginable as compared to other technologies that the world has experienced. It is a holistic technology that has demonstrated its efficiency in the present world through the industries that have adopted it to enhance their operations. An exploration into the plethora of benefits that biometric technology offers to the people is a pathway into understanding the importance of technological advancement in the modern world and the posterity. I believe this is a critical part because it cuts across controversy and criticism from millions of people that only hear or read about this technology without realizing the manner in which the world can change with its full adoption.

## VI. CONCLUSION

When it relates to what biometric is actually about, the use of biometrics in our daily and professional life looks routine yet incredibly careless. Based on results, it has been established that biometrics, which have offered a distinctive method of identification and verification for the privacy and security of information are challenging to cross. The integration of machine learning and artificial intelligence with biometrics is considered to be the direction of biometric technology in the future. The purpose of this study is to examine the developments in biometric systems from an engineering and medical perspective. The study looked at data collecting from various angles, including what biometrics is, what surrounding the technology, and what it might be used for in the future. According to qualitative data analysis, the findings of the respondent response reveal that, although using biometrics in its most basic form on a regular basis, the majority of respondents (especially those from the medical sector) have limited knowledge of the field. This merely serves to demonstrate the idea that many people rely more on technology because it is convenient and practical than because they are aware of its benefits and drawbacks. The participants from the healthcare and engineering sectors shared a common knowledge of the fundamental security and characteristics of a person of biometric systems, whereas the attendees from the engineering and medical sectors were more aware of the risk that biometric technology may face to human existence.

The results of this study show that, when properly investigated and included, biometric technology has the potential to flourish. Although the field of biometrics offers bright futures, its ability to grow has been constrained by a lack of knowledge and opportunities in the medical and engineering fields. Based on information gathered from the engineering and health care fields as a whole, this study examines and forecasts the use of biometric systems. As a follow-up to this study, additional research into biometric techniques is planned.

## VII. FUTURE WORK

Data security and identification techniques have improved as a result of biometric technologies. There are still several areas of biometric technology research that need to be filled, though. For upcoming research on biometric systems, the suggestions listed below have been made.

- It is recommended that biometrics courses be included in school curricula to help students fully appreciate the benefits of the technology and to raise awareness of it.
- A study to be done on biometric technologies in workplaces and classrooms for identification purposes, adaptation to biometric systems should be accelerated.
- A study to be done on the importance of biometric research and technology so that specialists can better grasp the field's application.